\documentclass[lettersize,journal]{IEEEtran}
\usepackage{amsmath,amsfonts}
\usepackage{algorithmic}
\usepackage{algorithm}
\usepackage{array}
\usepackage[caption=false,font=normalsize,labelfont=sf,textfont=sf]{subfig}
\usepackage{textcomp}
\usepackage{stfloats}
\usepackage{url}
\usepackage{verbatim}
\usepackage{graphicx}
\usepackage{cite}
\usepackage{ulem}
\usepackage{xcolor}
\usepackage{tikz}  
\usepackage[mathscr]{eucal}
\usepackage{amsmath}
\usepackage{color}
\usepackage{multirow}
\usepackage{booktabs}
\usepackage{bm}
\begin{document}

\title{Object-Attribute-Relation Representation Based \\ Video Semantic Communication}

\author{
  Qiyuan Du, ~\IEEEmembership{Student Member,~IEEE,} 
  Yiping Duan, ~\IEEEmembership{Senior Member,~IEEE,}
  Qianqian Yang, ~\IEEEmembership{Member,~IEEE,} \\
  Xiaoming Tao, ~\IEEEmembership{Senior Member,~IEEE,}
  and M\'{e}rouane Debbah, ~\IEEEmembership{Fellow,~IEEE,}
  
\thanks{This work was supported by State Key Laboratory of Space Network and Communications, the National Natural Science Foundation of China (Nos. NSFC 62442106, 62227801, 62322109, 62171257 and U22B2001), New Cornerstone Science Foundation through the XPLORER PRIZE, and the Tsinghua University (Department of Electronic Engineering)-Nantong Research Institute for Advanced Communication Technologies Joint Research Center for Space, Air, Ground and Sea Cooperative Communication Network Technology. \textit{(Corresponding author: Xiaoming Tao.)}}
\thanks{
Qiyuan Du, Yiping Duan and Xiaoming Tao are with the Department of Electronic Engineering, Tsinghua University, Beijing, China, Beijing National Research Center for Information Science and Technology (BNRist), Beijing, China, and also with the State Key Laboratory of Space Network and Communications, Beijing, China (email: dqy21@mails.tsinghua.edu.cn, yipingduan@mail.tsinghua.edu.cn, taoxm@tsinghua.edu.cn). Xiaoming Tao is also with School of Computer Science and Technology, Xinjiang University, Xinjiang, China.
}
\thanks{
Qianqian Yang is with College of Information Science and Electronic Engineering, Zhejiang University, Hangzhou, China (email: qianqianyang20@zju.edu.cn).
}
\thanks{
M\'{e}rouane Debbah is with KU 6G Research Center, Department of Computer and Information Engineering, Khalifa University, Abu Dhabi 127788, UAE (email: merouane.debbah@ku.ac.ae) and also with CentraleSupelec, University Paris-Saclay, 91192 Gif-sur-Yvette, France.
}
}





\maketitle
\pagestyle{empty}
\thispagestyle{empty}
\begin{abstract}
  With the rapid growth of multimedia data volume, there is an increasing need for efficient video transmission in applications such as virtual reality and future video streaming services. Semantic communication is emerging as a vital technique for ensuring efficient and reliable transmission in low-bandwidth, high-noise settings. However, most current approaches focus on joint source-channel coding (JSCC) that depends on end-to-end training. These methods often lack an interpretable semantic representation and struggle with adaptability to various downstream tasks. In this paper, we introduce the use of object-attribute-relation (OAR) as a semantic framework for videos to facilitate low bit-rate coding and enhance the JSCC process for more effective video transmission. We utilize OAR sequences for both low bit-rate representation and generative video reconstruction. Additionally, we incorporate OAR into the image JSCC model to prioritize communication resources for areas more critical to downstream tasks. Our experiments on traffic surveillance video datasets assess the effectiveness of our approach in terms of video transmission performance. The empirical findings demonstrate that our OAR-based video coding method not only outperforms H.265 coding at lower bit-rates but also synergizes with JSCC to deliver robust and efficient video transmission.
\end{abstract}


\section{Introduction}

\begin{figure}
  \centering
  \includegraphics[width=0.48\textwidth]{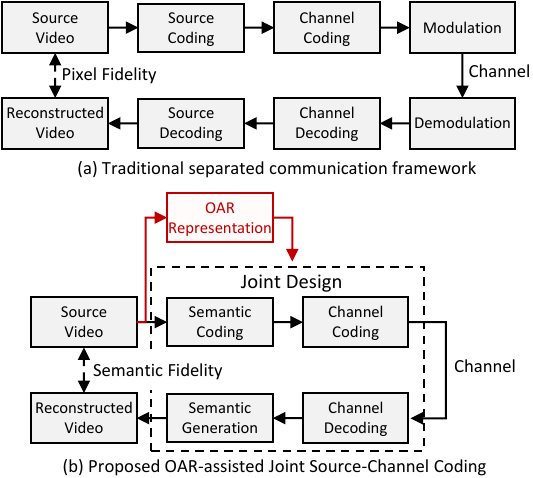}
  \caption{Comparison between traditional communication, classical semantic communication (without the OAR representation module) and the proposed OAR-based video transmission frameworks.}
  \label{fig: framework comparison}
\end{figure}

\IEEEPARstart{A}{s} video services increasingly dominate the multimedia traffic, the need for low bit-rate video compression and efficient transmission is becoming more critical. Traditional methods typically use a separate source-channel coding strategy to ensure pixel and symbol accuracy, as illustrated in Figure~\ref{fig: framework comparison} (a). However, with diminishing returns from standard video encoders\cite{DCVC}, the video quality at low bit-rates fails to satisfy user perception and task-specific requirements. Moreover, the distinct separation of source and channel coding can jeopardize the reliable transmission of video, especially over channels with low signal-to-noise ratios (SNRs), leading to increased instances of decoding failures\cite{ImageJSCC}.

Semantic communication\cite{SemanticCommunication} is emerging as a pivotal framework capable of surpassing the Shannon limit. By focusing on the semantic level of data characterization and transmission, it moves beyond traditional bit and symbol constraints, enhancing task-specific communication performance\cite{SemanticCom}. With the support of deep learning technologies\cite{DeepSemantic}, semantic communication has advanced significantly in both theory and practice, outperforming conventional communication methods in various applications\cite{TextSemantic, SpeechSemantic, RLASC, VideoJSCC, han2022semantic}.

Semantic information extraction and joint source-channel coding (JSCC) are crucial for semantic communication systems aiming for efficient transmission. By enabling effective semantic representations, these systems can prioritize critical semantic information (e.g., by allocating more bits or transmitted symbols) to enhance transmission efficiency. Recent advancements in image coding using semantic representations have shown promising results\cite{RLASC,Sketch}. Additionally, leveraging deep learning, deep JSCC methods have been developed and shown to outperform traditional separate coding frameworks in transmission performance\cite{ImageJSCC, DynamicJSCC, OFDMJSCC, DeepWive, PredictJSCC}. Notably, for structured multimedia such as audio and video, JSCC methods have achieved efficient transmission even at low signal-to-noise ratios (SNRs), significantly reducing the time and bandwidth costs caused by decoding failures. An overview of the JSCC framework is depicted in Figure~\ref{fig: framework comparison} (b).

Research into explicit semantic representation and transmission of videos is still limited. Studies such as \cite{VR+HD, SegmentationRepresentation, AG, vid2vid, AG2Video} have explored using semantic segmentation maps and scene graphs for semantic representations, implementing corresponding image and video generation to preserve the videos' semantic content.
Deep learning methods such as object detection\cite{FasterRCNN, YOLO}, classification\cite{Resnet} and relation recognition\cite{RelationDetection2, VisualGenome, RelationDetection} have been employed for scene semantic parsing and semantic organization.
However, these studies did not deeply engage with compression perspectives or data volume reduction. In \cite{Danlan2021, DSSLIC, RLASC}, semantic segmentation maps served as explicit semantic representations for image compression, enhancing semantic fidelity by varying bit-rates across different categories. These efforts treated semantic information as supplementary to traditional image compression, relying still on transmitting numerous pixel and residual features for quality enhancements, which hindered significant bit-rate reductions. \cite{diffusionSC} took a different approach by encoding and transmitting only semantic segmentation maps and reconstructing images using a diffusion model, catering to extreme channel conditions. Similarly, \cite{Sketch} used sketch representations and generative models. In video conferencing, efficient transmission was achieved using facial keypoints\cite{VideoConference}. Nevertheless, these approaches primarily focused on the semantic representation and coding of images or simpler video formats. The challenge remains to explore these techniques for large-scale, time-continuous, and object-consistent video representation to fully address the complexities of large scene videos.

Several JSCC approaches have been proposed for video transmission, yet there remains a dearth of research delving into explicit semantic representation and extraction of video contents. Approaches like DeepWive \cite{DeepWive} and \cite{Zhang2023} have achieved residual estimation based video transmission by employing bidirectional optical flow estimation, along with channel bandwidth allocation among frames. DVST \cite{VideoJSCC} and MDVSC \cite{MDVSC} have utilized nonlinear transforms and shared prior distributions for channel bandwidth allocation in features. However, these endeavors have predominantly focused on implicit latent features which are usually outputs of some neural layers. While ABRVSC \cite{ABRVSC} instead transmits category labels of videos frames for video sensing tasks, the receiver however can only recover category labels, resulting in a loss of significant video contents when adopted for video reconstruction. Thus, reliable video transmission with explicit semantic representation and extraction needs to be further explored.

To address the above issues, this paper proposes a novel video transmission system leveraging object-attribute-relation (OAR) representation model of semantic information, as shown in Fig.~\ref{fig: framework comparison} (b). Specifically, the OAR model characterizes scene semantics by capturing key objects, their attributes, and the relationships within video frames, tailored to the requirements of downstream applications. Benefiting from advancements in existing deep learning algorithms\cite{FasterRCNN,YOLO,Resnet,RelationDetection2}, OAR-based semantic information can be efficiently and automatically extracted.
Compared with object edges and semantic segmentation maps widely used in previous studies, OAR captures more structured semantic correlations, which are represented by connected graphs; compared with scene graph representation, OAR pays more attention to the enrichment of attributes, which enables better reconstruction of videos. In addition, this paper implements a JSCC pipeline for video transmission by exploiting the advantages of compact and downstream tasks related OAR data. Specifically, this paper implements OAR-assisted JSCC for key frames so that the model can pay more attention to the key objects in the scene. Experimental validation is performed on traffic surveillance videos, and the results indicate that the proposed system can still maintain better foreground reconstruction quality under low bandwidth and SNR conditions.

In summary, the contributions of this paper include the following:

\begin{enumerate}
  \item We propose an OAR-based video representation and generative reconstruction system for video transmission. In particular, OAR graphs characterize the motion and state changes of objects in a video, enabling low bit-rate video coding. This preserves foreground objects and semantics even at low bit-rates, thus overcoming the performance degradation of traditional video encoders for downstream tasks.
  \item We propose an OAR-assisted JSCC method for the transmission of key video frames. This method uses OAR as the side information to the channel coding and image reconstruction process, which leads to the enhancement of perceptual quality; For non-key video frames, we transmit only OAR representations. 
  \item Experiments on a traffic surveillance dataset show that the proposed method improves the perceptual quality of reconstructed videos significantly compared to conventional methods such as H.265 and prior work in semantic communication, namely DVST. Specifically, at a CBR of 1/300, the proposed method exhibits an average reduction in LPIPS loss of 0.054 when compared to DVST. Additionally, it facilitates improved performance in object detection tasks, achieving an average increase in mean average precision (mAP) of 0.15 compared to DVST.
\end{enumerate}

The remainder of this paper is organized as follows: in section~\ref{section 2}, we introduce the OAR-based video coding and reconstruction method. In section~\ref{section 3}, OAR-based image JSCC system is introduced, and a video transmission system that combines OAR representation and image JSCC is presented. Section~\ref{section 4} includes the experimental design and the results of the OAR-based video coding method and the transmission system. Section~\ref{section 5} includes the analysis, discussion and ablation studies. Section~\ref{section 6} summarizes the entire paper.

\section{Object-Attribute-Relation-Based Video Representation and Coding}
\label{section 2}

\begin{figure*}
  \centering
  \includegraphics[width=1.0\textwidth]{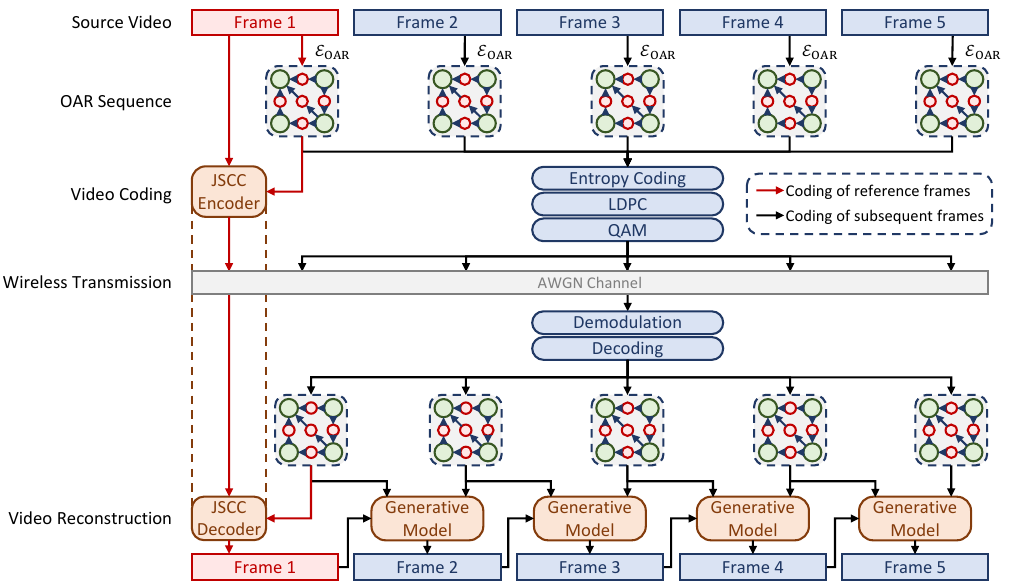}
  \caption{Overall framework of OAR-based video compressive coding and transmission. Frames are represented by OAR and transmitted via LDPC channel coding and QAM modulation. Additionally, the reference frame is coded and transmitted via OAR-assisted JSCC.}
  \label{fig: overall system}
\end{figure*}

\subsection{Object-Attribute-Relation Formulation and OAR Based Video Communication Framework}

An Object-Attribute-Relation (OAR) representation is a structured abstraction of the scene semantics in the form of a graph. For image $\bf I$, the OAR is expressed as a tuple of objects, attributes and relations: ${\bf I}\stackrel{\mathcal{E}_{OAR}}{\longrightarrow}\left(\mathcal{O},\mathcal{A},\mathcal{R}\right)$, 
where $\mathcal{E}_{OAR}$ denotes the mapping from the image to OAR. For a video with $T$ consecutive frames, the OAR is represented as a temporal sequence, $[\left(\mathcal{O}_1,\mathcal{A}_1,\mathcal{R}_1\right),\cdots,\left(\mathcal{O}_T,\mathcal{A}_T,\mathcal{R}_T\right)]$.

As shown in the expression, OAR contains three components: ${\mathcal{O}}=[o_1,\cdots,o_N]$, ${\mathcal{A}}=[{\bf a}_1,\cdots,{\bf a}_N]$, and ${\mathcal{R}}=\{r_1,\cdots,r_K\}$, where $N, K$ denote the numbers of objects and relations, respectively.
${\mathcal{O}}$ is the object set with $o_n$ denoting a unique ID for each object. ${\mathcal{A}}$ is the attribute set, with ${\bf a}_n=\left(a^{(n)}_1,a^{(n)}_2,\cdots,a^{(n)}_A\right)={\mathcal{E}_\text{A}}({\bf I},o_n)$ denoting $A$ attributes of object $o_n$. 
The relation is denoted by ${\mathcal{R}}=\left\{{\bf r}_1,\cdots{\bf r}_R\right\}=\left\{(o_{\text{s},1},o_{\text{o},1},r_1),\cdots,(o_{\text{s},K},o_{\text{o},K},r_K)\right\}$, $r_k=r_{o_{\text{s},k}\rightarrow o_{\text{s},k}}$ representing the relation with subject $o_{\text{s},k}$ and object $o_{\text{o},k}$.
An OAR is characterized by a directed graph with objects as nodes and relations as directed edges.
The relations and attributes (e.g., category, color, etc.) take values from a deterministic set, i.e.,
\begin{equation}
  \begin{aligned}
  a_p^{(n)}&\in\mathscr{A}_p,p=1,2,\cdots,A \\
  r_{o_i\rightarrow o_j} &\in \mathscr{R}, i,j=0,1,2,\cdots,N.
  \end{aligned}
\end{equation}

OAR is highly interpretable and has a small data volume because only semantics of key objects are characterized.
It is especially conducive to task-oriented applications with steady backgrounds, such as surveillance videos.
Additionally, OAR provides scalability and can be adapted to complex scenarios through the extension of attributes and relations.
Based on the above modeling, this article proposes an OAR-based video transmission framework, as presented in Fig.~\ref{fig: overall system}.

In the proposed framework, $T$ consecutive frames are encapsulated as one group-of-pictures (GoP). The first frame of each GoP (known as the reference frame or key frame) is intra-coded and transmitted, providing contour and texture information about the background and key objects. The semantics of all frames in a GoP are represented as OARs and transmitted after entropy coding, channel coding and modulation. OAR sequences are decoded after demodulation at the receiver, and generative reconstruction is carried out based on the reference frame incorporated with OAR sequences.
Without loss of generality, each GOP is independently processed. The reference frames are denoted by ${\bf I}_1$, and subsequent $T-1$ frames are denoted by ${\bf I}_t, t=2,\cdots,T$.
For objects contained in reference frames, faithful reconstruction is realized based on the motion information in OAR. Newly appeared objects are generated based on the prior knowledge of the trained models,
resulting in possible bias in pixel-level details such as colors and textures. This article focuses on task-oriented video transmission, so deviations in detailed textures are acceptable under the premise of semantic fidelity.

\subsection{OAR Extraction and Representation}

OAR representation is the critical component of the proposed video transmission system, providing semantics with low bit-rate representation and enabling efficient video transmission.Semantic concepts require specifications depending on different scenarios and tasks.
Consider a surveillance video transmission system oriented toward tasks such as object detection and anomaly monitoring. Vehicles in the foreground are treated as the objects in OAR. The attributes consist of the position ${\bf p}=(x,y)$, size ${\bf s}=(w,h)$, angle $\theta$, and category $c$, denoted by ${\bf a}=(a_1,a_2, a_3,a_4)=({\bf p},{\bf s},\theta,c)$. 
Relations involve mainly ``occlusions" between foreground objects and occlusions of the foreground by the background. The trivial relation of objects ``in" the background is also included.
Consequently, for an image of size $W\times H$, the attributes and relations take the value space of:
\begin{equation}
\begin{array}{rll}
{\bf p},{\bf s}&\in\mathscr{A}_1&=\mathscr{A}_2=[0,W]\times[0,H] \\
\theta &\in\mathscr{A}_3&=[0,360) \\
c&\in\mathscr{A}_4&=\{\text{car,bus,van,others,background}\} \\
r_{o_i\rightarrow o_j}&\in \mathscr{R}&=\{\text{occlusion,in,null}\}.
\end{array}
\end{equation}

To realize automated OAR extraction, multi-object tracking (MOT) is utilized, followed by several attribute recognition models. The overall process consists of three parts: multi-object tracking, attribute recognition, and relation identification.

First, multi-object tracking. Object detection is first conducted via YOLOv9\cite{yolov9} to obtain the categories $c_n^{(t)}$ and bounding boxes ${\bf b}_n^{(t)}=(x_n^{(t)},y_n^{(t)},w_n^{(t)},h_n^{(t)})$, where $n=1,\cdots,N_t$. To match the identical objects in consecutive frames, a unique ID $o_n^{(t)}$ is assigned to each object via the SORT algorithm\cite{SORT}. IDs for all objects constitute the object component ${\mathcal{O}_t}=[o_1^{(t)},\cdots,o_{N_t}^{(t)}]$ in OAR.

Second, attribute recognition. The object pose, i.e., orientation $\theta_n^{(t)}$ in the 2D pixel plane, is estimated via an angle estimation network EgoNet \cite{EgoNet}.
Combined with the category and bounding box, they collectively constitute the attribute component in the OAR: ${\mathcal{A}_t}=[{\bf a}_1^{(t)},\cdots, {\bf a}_{N_t}^{(t)}]$, where ${\bf a}_n^{(t)}=({\bf p}_n^{(t)},{\bf s}_n^{(t)},\theta_n^{(t)},c_n^{(t)})$.

Third, relation identification. 
Object pairs with intersecting bounding boxes are examined and the occlusion relations are determined by comparing the appearance of the overlapping regions with the respective object features.
In addition, occlusions between objects and background elements are also considered. Regions located at the spatial forefront are manually labeled. Objects overlapping with the labeled area are considered to be occluded by the background. 
By incorporating the identified ``occlusion" and the trivial ``in" relation, the relation component in OAR is formulated as $\mathcal{R}_t=\{{\bf r}_1^{(t)},\cdots,{\bf r}_K^{(t)}\}$.

\begin{figure*}
  \centering
  \includegraphics[width=1.0\textwidth]{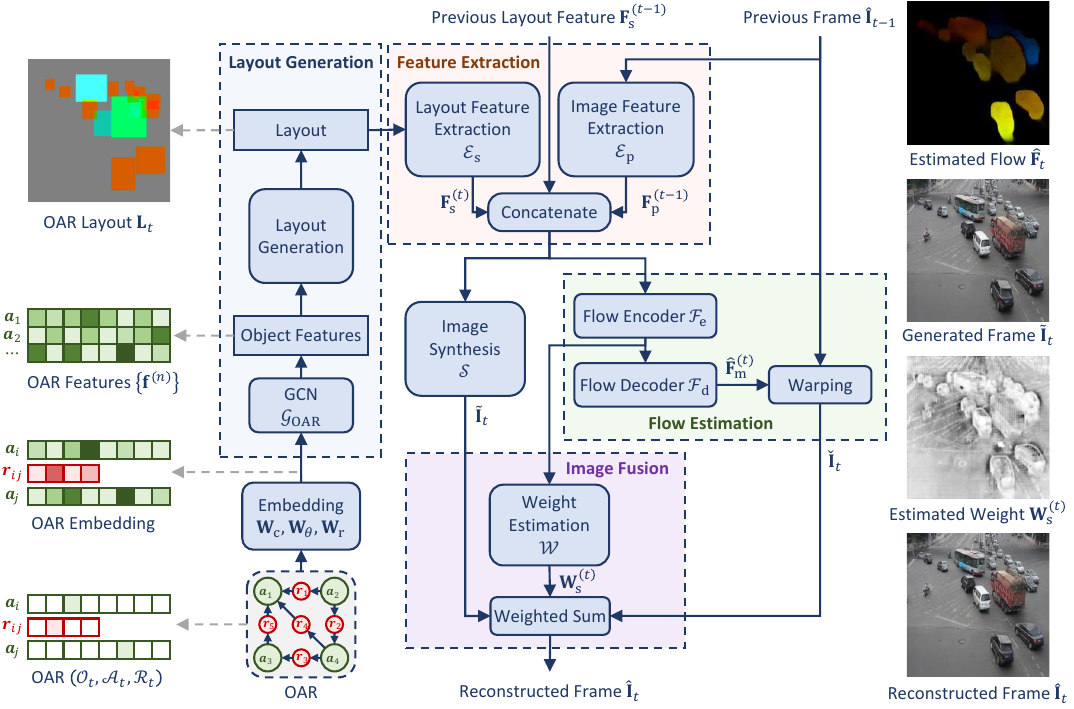}
  \caption{Framework of OAR-based video generation and visualizations for intermediate results.}
  \label{fig: generation}
\end{figure*}

\subsection{OAR Based Video Generative Reconstruction}
\label{reconstruction}
At the receiver, generative reconstruction is conducted based on the received OAR sequence combined with reference frames. The overall framework is shown in Fig.~\ref{fig: generation}, which can be organized into four modules: (1) OAR embedding and graph computing, (2) OAR layout generation, (3) optical flow estimation and frame deformation, and (4) image synthesis and fusion. Because the OAR for each video frame is processed via the same process of OAR embedding, graph computing and layout generation, the time superscripts are omitted in modules (1) and (2) for brevity.

\subsubsection{OAR Embedding and Graph Computing}

The category $c_n$, angle $\theta_n$ (quantized by $q$-bits) and relation $r_{k}$ are first represented by one-hot vectors ${\bf c}_n,\bm{\theta}_n,{\bf r}_k$. Three projection matrices ${\bf W}_c,{\bf W}_\theta,{\bf W}_r$ are learned to map the vectors into an embedding space \cite{Word2Vec}, denoted by:
\begin{equation}
\begin{aligned}
{\bf e}_\text{c}^{(n)}&={\bf W}_\text{c}{\bf c} \\
{\bf e}_\theta^{(n)}&={\bf W}_\theta\bm{\theta} \\
{\bf e}_\text{r}^{(k)}&={\bf W}_\text{r}{\bf r}.
\end{aligned}
\end{equation}

For the object ``background", the category is embedded via the same method, and the angle embedding vector is set to a constant ${\bf 0}$.

Although OAR realizes the extraction of attributes and relations between objects in the scene, there are still close connections and interactions between objects. This cross attention across space and individuals inspires us to perform propagation and modulation of features across objects after embedding, and the relations are an important basis for this process. Considering graph convolutional networks (GCNs) have been utilized to extract high-level semantic features from graph-structured data and have shown excellent performance on classification tasks \cite{GCN,gcn4text} and generative tasks \cite{sg2image,sg2image2} of graph data,
a GCN (denoted by $\mathcal{G}_\text{OAR}$) is utilized with reference to \cite{GCN} for deep feature extraction of OARs. Specifically, a primary graph is constructed with the concatenation of category and angle embedding vectors ${\bf e}^{(n)}=\text{concat}({\bf e}_c^{(n)},{\bf e}_\theta^{(n)})$ as nodes and the relation embedding vectors ${\bf e}_\text{r}^{(k)}$ as directed edges from $o_{\text{s},k}$ to $o_{\text{o},k}$. The primary graph is subsequently processed by a two-layer GCN to obtain a deep feature for each object, denoted by ${\bf f}^{(n)}$.
Multilayer GCNs further expand the scope of attention so that the obtained features can be affected by the associated multiple objects simultaneously.

\subsubsection{OAR Layout Generation}

The temporal dynamics of the motion and scale evolution of objects are captured by the location and size attributes of the OAR. 
With reference to AG2Video\cite{AG2Video}, the feature vector of each object is propagated to obtain the layout feature ${\bf L}\in\mathbb{R}^{H\times W\times D}$. The element ${\bf L}_{i,j,d}$ is formulated as:
\begin{equation}
{\bf L}_{i,j,d}=\sum_{n=1}^{N}{\bf L}^{(n)}_{i,j,d},
\end{equation}

where ${\bf L}^{(n)}\in\mathbb{R}^{H\times W\times D}$ represents the individual layout of the $n$-th object. Specifically, the deep feature ${\bf f}^{(n)}$ is expanded to the region corresponding to the bounding box ${\bf b}_n=(x_n,y_n,w_n,h_n)$, which is computed as follows:
\begin{equation}
{\bf L}^{(n)}_{i,j,d}=
\left\{
\begin{array}{ll}
\begin{aligned}
&{\bf f}_d^{(n)}, & & \text{if } y^{(n)} \le i \le y^{(n)}+h^{(n)}, \\
& & & \text{and } x^{(n)} \le j \le x^{(n)}+w^{(n)}, \\
&0, & & \text{otherwise}.
\end{aligned}
\end{array}\right.
\end{equation}

One layout is generated from the OAR of each frame, constructing the layout sequence: $[{\bf L}_1,\cdots,{\bf L}_T]$.

\subsubsection{Optical Flow Estimation and Frame Deformation}

With reference frames providing appearance information of objects, optical flows are estimated based on the motion information of OAR and frame deformation is conducted. The $t$-th frame ($t\ge2$) is predicted based on the previously synthesized frame $\hat{\bf I}_{t-1}$ and adjacent layouts ${\bf L}_{t-1}, {\bf L}_t$. First, a semantic feature extractor $\mathcal{E}_\text{s}$ and a pixel feature extractor $\mathcal{E}_\text{p}$ are utilized to obtain deep OAR features of the two frames ${\bf F}_\text{s}^{(t)}=\mathcal{E}_\text{s}({\bf L}_t), {\bf F}_\text{s}^{(t-1)}=\mathcal{E}_\text{s}({\bf L}_{t-1})$, and image features of the previous frame ${\bf F}_\text{p}^{(t-1)}=\mathcal{E}_\text{p}(\hat{\bf I}_{t-1})$, respectively. Optical flows $\hat{\bf F}_\text{m}^{(t)}\in\mathbb{R}^{H\times W\times 2}$ are then estimated via an optical flow estimator $\mathcal{F}$ in two stages:
\begin{equation}
\begin{aligned}
\hat{\bf F}_\text{m}^{(t)}&=\mathcal{F}({\bf F}_\text{s}^{(t)},{\bf F}_\text{s}^{(t-1)},{\bf F}_\text{p}^{(t-1)}) \\
&=\mathcal{F}_{\text{d}}(\mathcal{F}_{\text{e}}({\bf F}_\text{s}^{(t)},{\bf F}_\text{s}^{(t-1)},{\bf F}_\text{p}^{(t-1)})).
\end{aligned}
\end{equation}

Based on the estimated optical flows $\hat{\bf F}_\text{m}^{(t)}$, the predicted image can be obtained by $\check{\bf I}_{t}=\text{warp}(\hat{\bf I}_{t-1},\hat{\bf F}_\text{m}^{(t)})$.

\begin{figure*}
  \centering
  \includegraphics[width=1.0\textwidth]{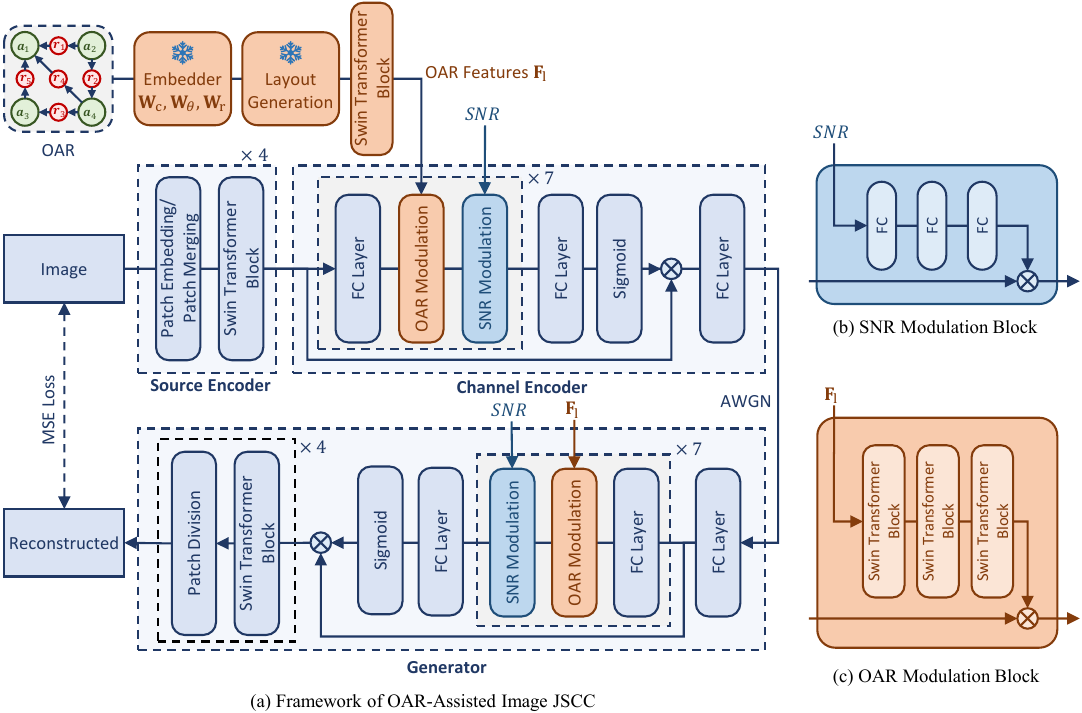}
  \caption{The framework of OAR-assisted image JSCC. The SNR is assumed to be available at both the transmitter and the receiver. OAR undergoes lossless transmission, ensuring that both the transmitter and receiver obtain identical OAR features ${\bf F}_\text{l}$ through the utilization of OAR feature extraction networks with identical parameters.}
  \label{fig: OAR JSCC}
\end{figure*}

\subsubsection{Image Synthesis and Fusion}

While frame deformation enables motion reconstitution, occluded regions and newly emerging objects cannot be resolved. Accordingly, generative reconstruction is employed for detail complementation with reference to vid2vid \cite{vid2vid}. Specifically, an image synthesis network $\mathcal{S}$ is implemented to generate the current frame
by
\begin{equation}
\tilde{{\bf I}}_t=\mathcal{S}({\bf F}_\text{s}^{(t)},{\bf F}_\text{s}^{(t-1)},{\bf F}_\text{p}^{(t-1)}).
\end{equation}

Finally, a weight estimation network $\mathcal{W}$ is implemented for mask prediction from optical flow features, denoted by ${\bf W}_\text{s}^{(t)}=\mathcal{W}(\mathcal{F}_{\text{e}}({\bf F}_\text{s}^{(t)},{\bf F}_\text{s}^{(t-1)},{\bf F}_\text{p}^{(t-1)}))\in[0,1]^{H\times W}$. The predicted image $\check{\bf I}_{t}$ and the synthesized image $\tilde{{\bf I}}_t$ are fused to obtain the final synthesis:
\begin{equation}
\hat{\bf I}_t={\bf W}_\text{s}^{(t)}\odot\tilde{\bf I}_{t}+\left(1-{\bf W}_\text{s}^{(t)}\right)\odot\check{{\bf I}}_t
\end{equation}

Denote the parameters of the network to be optimized as $F=\{{\bf W}_\text{c},{\bf W}_\theta,{\bf W}_\text{r},\mathcal{G}_\text{OAR},\mathcal{E}_\text{s},\mathcal{E}_\text{p},\mathcal{S},\mathcal{F}_\text{e},\mathcal{F}_\text{d},\mathcal{W}\}$, and the modules are jointly trained following \cite{vid2vid} by solving:

  \begin{equation}
\min_F\left(\max_{D_I}\mathcal{L}_I(F,D_I)+\max_{D_V}\mathcal{L}_V(F,D_V)\right)+\lambda_W\mathcal{L}_W(F), \label{optimization}
  \end{equation}

where subscripts $I$ and $V$ denote the optimization of single and consecutive video frames, respectively. $D_{I/V}$ denotes the discriminator, and $\mathcal{L}_{I/V}$ denotes the weighted summation of the VGG loss and the adversarial loss. $\mathcal{L}_W$ denotes optical flow loss, which guides the training process of the optical flow estimation module $\mathcal{F}_\text{e}\cdot\mathcal{F}_\text{d}$ in terms of the accuracy of the optical flow $\hat{\bf F}_\text{m}$ and the quality of the deformed image $\check{\bf I}_t$.

A set of intermediate results is shown in Fig.~\ref{fig: generation}. The layout ${\bf L}_t$, the estimated optical flows $\hat{\bf F}_\text{m}^{(t)}$, the image synthesis results $\tilde{{\bf I}}_t$, the estimated weights ${\bf W}_\text{s}^{(t)}$ and the final generated results $\hat{\bf I}_t$ are illustrated.

\section{OAR-assisted JSCC and Video Transmission}
\label{section 3}

In this section, video transmission over Gaussian channels is realized, including transmission of reference frames and OAR sequences. 
OAR-modulated JSCC is proposed for reference frame transmission with reference to SwinJSCC \cite{SwinJSCC}, and OAR sequences is transmitted via LDPC \cite{LDPC} and QAM.

\subsection{OAR-modulated JSCC}

For the coding and transmission of key frames, OAR representation is integrated with SwinJSCC\cite{SwinJSCC} as the backbone, which consists of a jointly trained source encoder $\mathcal{E}_\text{se}$, a channel encoder $\mathcal{E}_\text{ce}$ and a generator $\mathcal{G}$.
The overall framework is shown in Fig.~\ref{fig: OAR JSCC} (a), which contains two components: OAR feature extraction and OAR-assisted image JSCC.
Note that the backbone network and SNR modulation are derived from the original SwinJSCC\cite{SwinJSCC}. We introduce the proposed OAR modulation in the following.

The OAR layout ${\bf L}_1$ is first generated through the layout generation method implemented in section~\ref{reconstruction}. 
Deep layout features are subsequently obtained through a Swin Transformer (ST)\cite{Swin} block: ${\bf F}_\text{l}=\text{ST}_1({\bf L}_1)\in\mathbb{R}^{H/4\times W/4\times C}$.

In the backbone of the JSCC networks, the OAR modulation module is introduced into the channel encoder $\mathcal{E}_\text{ce}$ and generator $\mathcal{G}$, which is shown in Fig.~\ref{fig: OAR JSCC} (c). 
A set of multipliers ${\bf m}\in(0,1)^{h_\text{in}\times w_\text{in}\times 1}$ is learned from the layout feature ${\bf F}_\text{l}$ utilizing three consecutive ST blocks. The feature distribution is shifted
by performing a pointwise multiplication operation:
\begin{equation}
{\bf F}_\text{out}={\bf F}_\text{in}\odot{\bf m}.
\end{equation}

In conjunction with the Gaussian noise $\bf n$, the reconstructed image at the receiver can be denoted by:
\begin{equation}
    \hat{\bf I}_1=\mathcal{G}\left(\mathcal{E}_\text{ce}\left(\mathcal{E}_\text{se}({\bf I}_1)\mid{\bf L}_1 \right)+{\bf n}\mid{\bf L}_1\right).
\end{equation}

With OAR-based modulation, the JSCC system enhances object retention by prioritizing foreground regions. Moreover, the integration of OAR features into the generator offers additional semantic cues for the decoding of foreground objects, thereby improving adaptability to higher noise levels.

\subsection{OAR-based Video Transmission Pipeline}

This section integrates the OAR-based video coding system presented in section~\ref{section 2} with the OAR-modulated image JSCC to propose an OAR-based video transmission framework. 
As illustrated in Fig.~\ref{fig: overall system}, the framework comprises three components: JSCC of key frames, OAR transmission, and OAR-based video reconstruction. 
OAR-assisted source and channel coding are applied to key frames for transmission via Gaussian channels. Upon reception, reconstruction from the received symbols yields the reconstructed frame $\hat{\bf I}_1$, enabling the restoration of subsequent non-key frames.
OAR transmission is conducted via 1/3-rate LDPC and 4QAM modulation, which achieves distortion-free transmission. 
With the received reference frames and OAR sequences, the receiver performs video reconstruction based on the video generation network in \ref{reconstruction}.

\section{Experiments and Results}
\label{section 4}

\begin{figure*}
  \centering
  \includegraphics[width=1.0\textwidth]{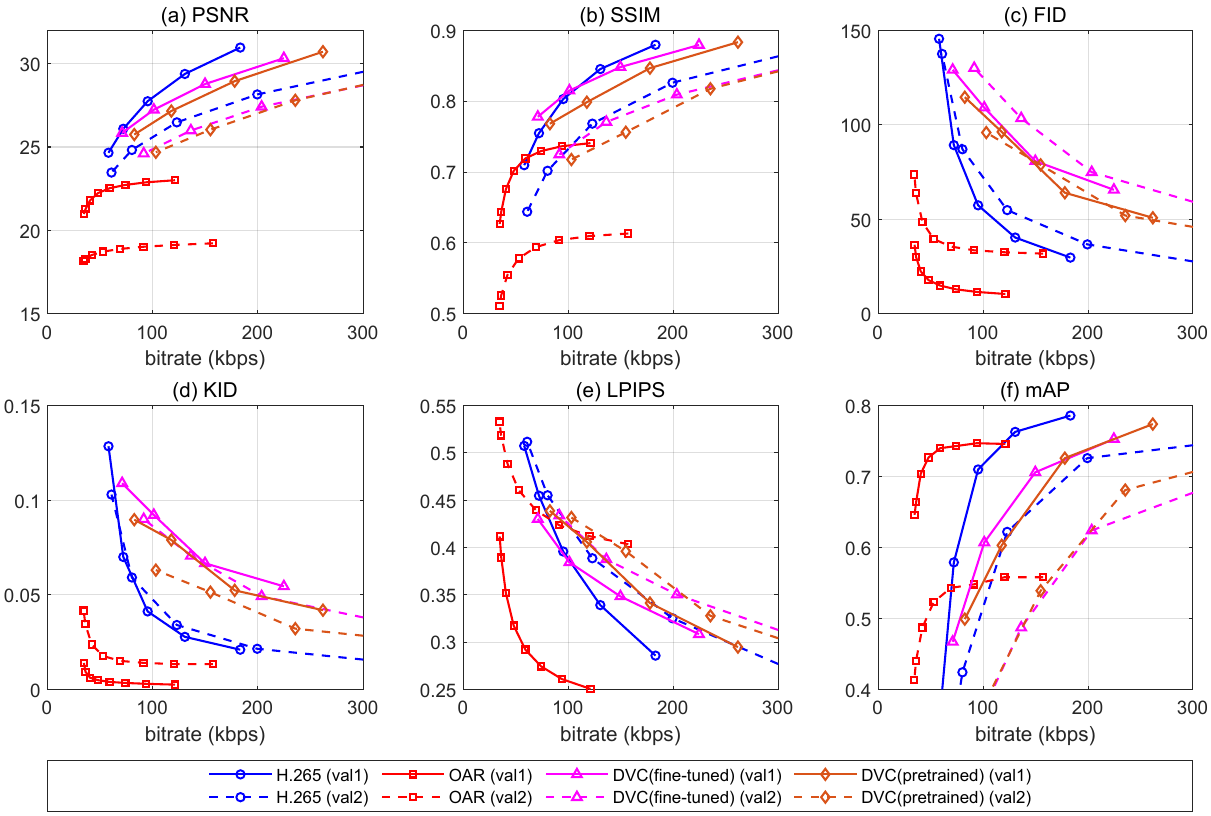}
  \caption{Performance comparison of the proposed OAR based method with H.264, H.265 and DVC at different bit-rates on the UA-DETRAC dataset.}
  \label{fig: RD}
\end{figure*}

\subsection{Dataset, Performance Metrics and Baselines}

\subsubsection{Dataset Preparation}

We conduct experiments mainly on the traffic surveillance video dataset UA-DETRAC\cite{DETRAC} for performance validation, and the videos are resized to $512\times512$.
Given the limitations of current MOT and attribute recognition algorithms in achieving accurate recognition, we utilize the ground truth for subsequent experimental analysis.
Additionally, we conducted experiments on the CATER\cite{CATER} and SoccerNet\cite{SoccerNet} datasets to demonstrate the generalizability of the proposed framework to other scenarios. The CATER dataset consists of videos of virtual scenes generated via 3D rendering. Objects such as spheres, squares, cones and cylinders of different sizes, colors and materials perform actions such as translations and rotations. The dataset is created for future immersive scene understanding, and experiments on this dataset help to validate the effectiveness of the proposed video transmission method for future immersive media and videos of interactive scenes. The SoccerNet dataset includes videos of several real soccer games, where the locations of players and balls are annotated. The videos of CATER and SoccerNet are resized to $256\times256$ and $512\times512$, respectively.

For dataset UA-DETRAC, video frames are segmented into a training set and two validation sets with a ratio of 8:1:1 for the number of frames.
The validation set \texttt{val1} shares the same scenes as the training set but features different objects and motions, facilitating the validation of generating diverse objects. This is adapted to scenarios where the model is deployed on a fixed camera and trained with historical data from that camera. Conversely, the validation set \texttt{val2} comprises scenes distinct from those of the training set, enabling an assessment of performance across different backgrounds. 
For the CATER and SoccerNet datasets, the training and validation sets are divided at the ratio of 4:1 and taken from videos of different scenes.

\subsubsection{Experiment Settings}

The models in the proposed framework are trained in two steps. First, the OAR-based video reconstruction model is trained with original reference frames. 
The optimization object shown in expression (\ref{optimization}) imposes constraints on reconstructed video quality across spatial and temporal dimensions as well as optical flow estimation accuracy.
The OAR-based JSCC network is distinct trained on AWGN channels. The symbol amplitudes are normalized to ensure an average signal power of 1, followed by the addition of AWGN noise sampled randomly at SNRs ranging from 0 to 20 dB to the transmitted symbols.

Experiments are conducted to validate the performance from two perspectives: video coding and wireless transmission. In both scenarios, video reconstruction is realized via the same model in \ref{reconstruction}. However, different compression and transmission methods are employed for reference frames. In the video coding task, reference frames undergo compression via BPG\cite{BPG} and are losslessly transmitted to evaluate the performance of the video reconstruction network. For wireless video transmission, reference frames are transmitted through OAR-assisted JSCC followed by an AWGN channel.

\begin{figure*}
  \centering
  \includegraphics[width=1.0\textwidth]{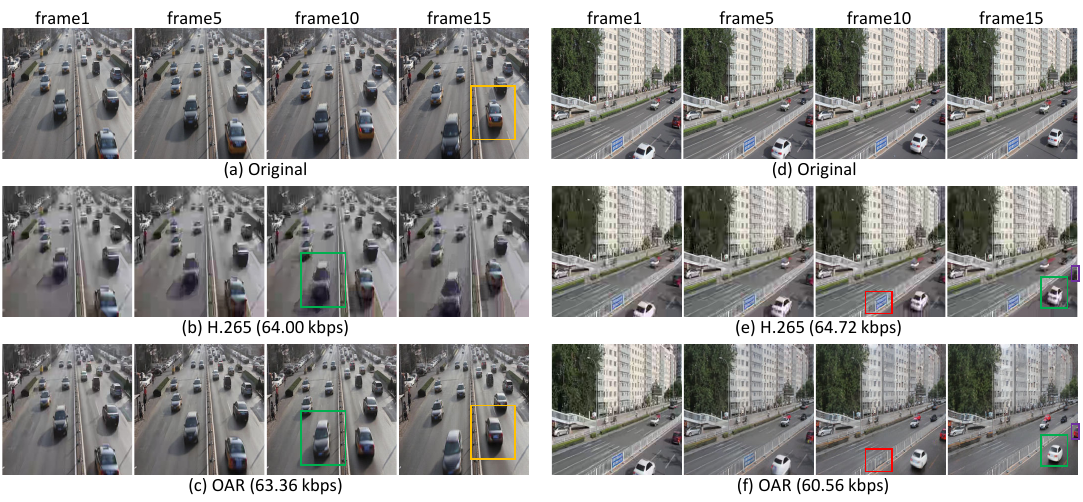}
  \caption{Visualization of the reconstruction results achieved by different algorithms. (a)-(c) correspond to the \texttt{val1} dataset and (d)-(f) correspond to the \texttt{val2} dataset. The bit-rates achieved by H.265 and the proposed OAR-based methods are labeled.}
  \label{fig: gallary}
\end{figure*}

\subsubsection{Performance Metrics}

We evaluate performance from the perspectives of pixel fidelity, perceptual quality, distribution similarity, and task performance. Specifically, PSNR and SSIM assess pixel fidelity, LPIPS\cite{LPIPS} represents perceptual quality, and FID\cite{FID} and KID\cite{KID} measure distribution similarity. Additionally, mAP serves as a metric for task-level evaluation.

The PSNR and SSIM metrics, which are indicative of pixel-level similarity, are computed via TensorFlow. LPIPS, FID and KID metrics extract deep features
through neural networks, measuring feature similarity as the metric.
LPIPS and FID (KID) are calculated via open-source codes \cite{LPIPS} and \cite{fid-kid}, respectively.

In addition to the aforementioned metrics, this paper also conducts video quality assessment from a task-oriented perspective using the mean average precision (mAP) of object detection tasks. An object detection model is trained on the original video frames, and subsequently applied to perform object detection on the reconstructed videos. The mAP metric (mAP@.95) is then computed to reflect the video quality.
In the implementation of this paper, YOLOv5\cite{yolov5} is utilized after fine-tuning.

To assess transmission efficiency, three metrics are employed: the bit-rate (for lossless transmission), the channel bandwidth ratio (CBR, for wireless transmission) and the signal-to-noise ratio (SNR, for transmission through noisy channels). 
The bit-rate is calculated by dividing the total number of bits by the video duration.
For reference frames in \texttt{val1} and \texttt{val2} of UA-DETRAC, the average data amounts of OAR are 366 bits and 219 bits, respectively. Using a predictive coding approach, the average bit-rates of the OAR sequences are 3.5 kbps and 2.2 kbps, respectively.
For the CATER and SoccerNet datasets, the average bit-rate of the OAR sequences are 3.28 kbps and 4.49 kbps, respectively.

CBR is defined as the ratio of the number of transmitted symbols to the original number of source symbols. 
For dataset \texttt{val1} of UA-DETRAC, additional CBRs of $7.0\times10^{-4}$ and $2.7\times10^{-4}$ are expected for the OAR of the reference frame and the overall video, respectively.
For \texttt{val2}, the corresponding CBRs are $4.2\times10^{-4}$ and $1.7\times10^{-4}$. 
The average CBRs of OAR sequences for the CATER and SoccerNet dataset are $1.02\times10^{-3}$ and $1.40\times10^{-3}$, respectively. The variation in the average CBR of OAR sequences on different datasets is caused mainly by the resolution and object density.

The signal-to-noise ratio (SNR) is the ratio of the signal power $P_\text S$ to the noise power $P_\text N$. 
For an AWGN channel $N\sim G(0,\sigma^2)$, the SNR is expressed as $SNR(dB)=10\log_{10}P_\text{S}/\sigma^2$.
Without loss of generality, $P_\text S=1$ is adopted, i.e., both the emitting symbols of JSCC and the conventional method are power-normalized.

\subsubsection{Baselines and Comparison Algorithms}

For the proposed OAR-based video coding system, the performance is compared with conventional video codec H.265 and the deep learning-based codec DVC\cite{DVC}.
The H.265 codec utilizes the x265 implementation of ffmpeg\cite{ffmpeg}, and the bit-rate is adjusted by CRFs. DVC is fine-tuned
to achieve various bit-rates.
Both the pretrained model (labeled by pretrained) and the one fine-tuned on specific datasets (labeled by fine-tuned) are examined to mitigate the influence of dataset distribution.

For video transmission over AWGN channels, the experiments involve the transmission of reference frames and transmission of entire videos. For reference frames, the image undergoes encoding with BPG, followed by LDPC\cite{LDPC} coding, and modulation with either BPSK or QAM. BPG implementation is based on the open-source tool\cite{BPG}, and LDPC and BPSK/QAM modulations are adopted from \cite{LDPC_realize}.
Three parameter configurations of 1/3 rate (1536,4608) LDPC, 1/2 rate (3072,6144) LDPC and 2/3 rate (3072,4608) LDPC are examined for channel coding, and modulation schemes of BPSK, 4QAM, 16QAM and 64QAM are utilized for symbol mapping.
In addition, this paper compares the performance of the proposed OAR-based JSCC with the original model to assess the impact of OAR. Both JSCC models utilize identical training methods and network structures to ensure fairness in comparison. 

Furthermore, the performance of OAR-based video transmission over AWGN channels is compared with H.265 video coding followed by LDPC and QAM. In addition, the recent video JSCC method DVST\cite{VideoJSCC} is implemented and compared.
The DVST model is trained at SNR=10 dB and fine-tuned with different rate control parameters to achieve lower CBRs.

\begin{figure*}
  \centering
  \includegraphics[width=1.0\textwidth]{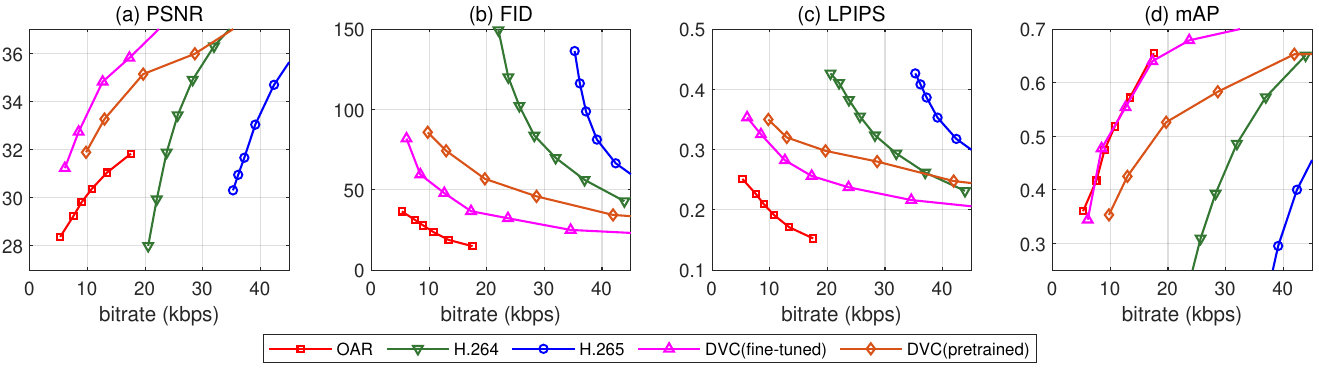}
  \caption{Performance comparison of the proposed OAR based method with H.264, H.265 and DVC at different bit-rates on the CATER dataset.}
  \label{fig: RD_CATER}
\end{figure*}

\begin{figure*}
  \centering
  \includegraphics[width=1.0\textwidth]{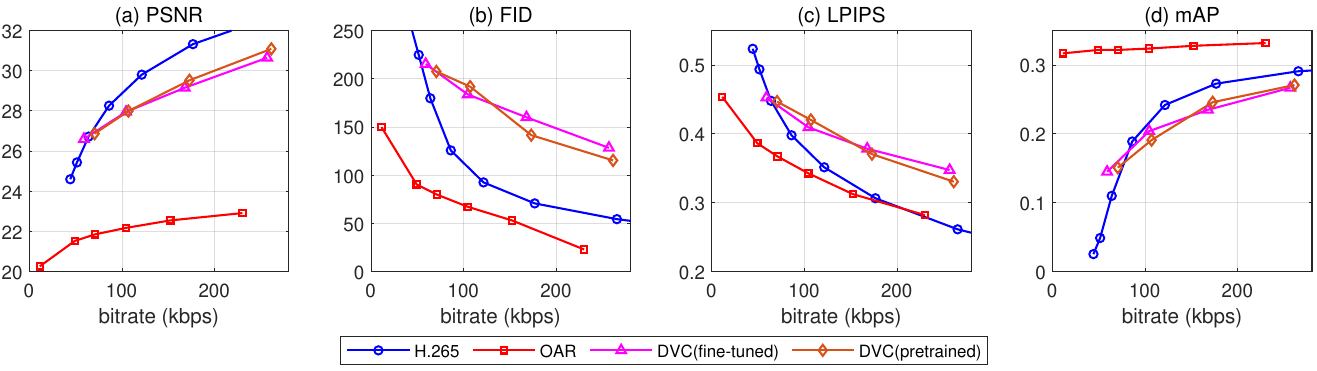}
  \caption{Performance comparison of the proposed OAR based method with H.265 and DVC at different bit-rates on the SoccerNet dataset.}
  \label{fig: RD_SoccerNet}
\end{figure*}

\subsection{Rate-Distortion Performance of OAR-Based Video Coding System}

Fig.~\ref{fig: RD} presents the performance comparison of the proposed OAR-based video coding framework with H.265 and DVC across different metrics on the UA-DETRAC dataset. The solid and dashed lines represent the performance on \texttt{val1} and \texttt{val2}, respectively. From Fig.~\ref{fig: RD} (a) and (b), it is evident that OAR-based video coding is inferior to H.265 and DVC in terms of pixel fidelity. This is because the proposed coding focuses primarily on downstream vision tasks, where detailed textures of the object and background are either provided by the reference frames or generated directly without relying on residuals. The discarding of prediction residuals plays a crucial role in achieving significant bit-rate reduction.

Despite the absence of residuals, the proposed method outperforms H.265 in terms of feature distribution and perceptual quality. As indicated in Fig.~\ref{fig: RD} (c) and (d), the proposed method better preserves the distribution properties in the feature dimension.
Conversely, H.265 exhibits significant feature bias due to the obvious block effect and blurring distortion at low bit-rates.
Similarly, according to the LPIPS metrics demonstrated in Fig.~\ref{fig: RD} (e), the proposed method achieves superior perceptual quality at lower bit-rates (especially below 100 kbps).

Furthermore, as depicted in Fig.~\ref{fig: RD} (f), the proposed method maintains high object detection performance as the bit-rate decreases, reflecting the effective reconstruction of foreground objects by OAR. Conversely, H.265 experiences significant performance degradation when the bit-rate falls below 100 kbps.
This is because the OAR contains vital object semantics, enabling better object recovery even at low bit-rates. Specifically, the position and category information within the OAR provides object motion and appearance features in the reconstructed video. From another perspective, the OAR-based video coding method achieves 60.56\% and 57.08\% bit-rate savings when the average mAP reaches 0.7 (\texttt{val1}) or 0.4 (\texttt{val2}), respectively. The performance of DVC is inferior to H.265 in general.

Fig.~\ref{fig: gallary} displays video frames reconstructed by H.265 and OAR-based coding at a bit-rate of approximately 65 kbps. As shown in the figure, the proposed method significantly reduces the block effect, enhancing the perceptual quality compared with H.265 (highlighted in the green boxes). Furthermore, objects at edges are better preserved because of the object position and category information in the OAR (marked by purple boxes). 

\begin{figure*}
  \centering
  \includegraphics[width=1.0\textwidth]{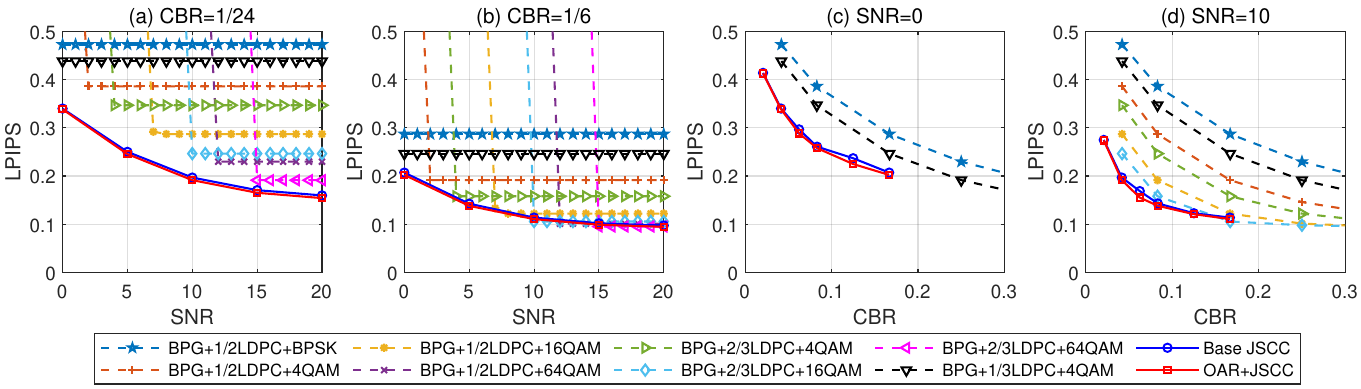}
  \caption{The image perceptual quality (LPIPS) at different CBRs and SNRs on the UA-DETRAC dataset. (a)-(b) illustrate the LPIPS versus SNR at different CBRs, and (c)-(d) illustrate the LPIPS versus CBR at different SNRs.}
  \label{fig: LPIPS-SNR_CBR}
\end{figure*}

\begin{figure*}
  \centering
  \includegraphics[width=1.0\textwidth]{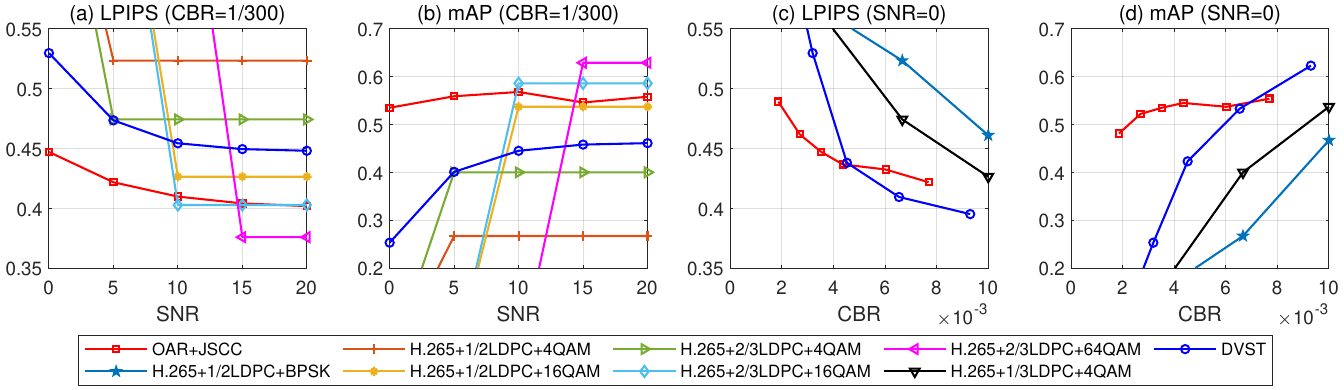}
  \caption{Performance comparison of OAR-based video transmission to H.265 paired with LDPC and DVST on the UA-DETRAC dataset. (a)-(b) are performances at CBR of 1/300, and (c)-(d) at SNR of 0 dB. The H.265 and H.264-based transmission under some parameters suffer from decoding failure and are not plotted. Same with Fig.~\ref{fig: video transmission_CATER} and Fig.~\ref{fig: video transmission_SoccerNet}}
  \label{fig: CBR-SNR-Distortion}
\end{figure*}

A comparison of \texttt{val1} and \texttt{val2} in Fig.~\ref{fig: RD} reveals that methods generally achieve lower performance on \texttt{val2}. This is primarily due to \texttt{val2} containing more complex motion patterns, posing challenges for inter-frame prediction and compensation. Moreover, \texttt{val2} contains more complex objects that are inherently harder to detect efficiently, leading to degraded object detection performance across all methods. For OAR-based coding, the distribution disparity between \texttt{val2} and the training set also results in performance degradation. As shown in Fig.~\ref{fig: gallary}, the object marked by the yellow box suffers from color deviation because the tail of the car is absent in the reference frame, whereas the background elements marked by the red box gradually fade. Despite this, in practical surveillance video applications, efficient adaptation through model fine-tuning is feasible, given the long processing period and relatively steady backgrounds. To ensure a fair performance comparison and mitigate the impact of the dataset distribution, subsequent analysis predominantly focuses on \texttt{val2}.

The video coding performances of the different methods on CATER and SoccerNet datasets are plotted in Fig.~\ref{fig: RD_CATER} and Fig.~\ref{fig: RD_SoccerNet}, respectively. For brevity, only four metrics, PSNR, FID, LPIPS and mAP, are presented. H.265 suffers from severe performance degradation due to more overheads on the low-resolution CATER dataset. Therefore, we also present the performance of H.264 to more fully demonstrate the advantages of the proposed method over conventional video coding frameworks. A comparison of Fig.~\ref{fig: RD_CATER} with Fig.~\ref{fig: RD} (val2) reveals that the proposed method achieves higher performance on the CATER dataset in all the metrics. This is because the CATER dataset has a relatively homogeneous background and the objects do not have complex textures, which is favorable for more fidelity video generative reconstruction. Fig.~\ref{fig: RD_CATER}(b) and (c) demonstrate that the proposed method achieves higher perceptual quality compared to H.265 and DVC, which is consistent with the experimental results on the UA-DETRAC dataset. In Fig.~\ref{fig: RD_CATER}(d), the fine-tuned DVC achieves comparable performance to the proposed method, which is due to the homogeneous distribution characteristics of the dataset.
As observed in Fig.~\ref{fig: RD_SoccerNet} (d), the performance of the proposed method in object detection on the SoccerNet dataset does not suffer from significant degradation with decreasing bit-rate. This results from the fact that there are only two categories in the dataset: person and ball. The generative model has a more explicit objective, and thus enables semantically consistent object generation despite the loss of texture information (at very low bit-rates) in the reference frame.

\begin{figure*}
  \centering
  \includegraphics[width=1.0\textwidth]{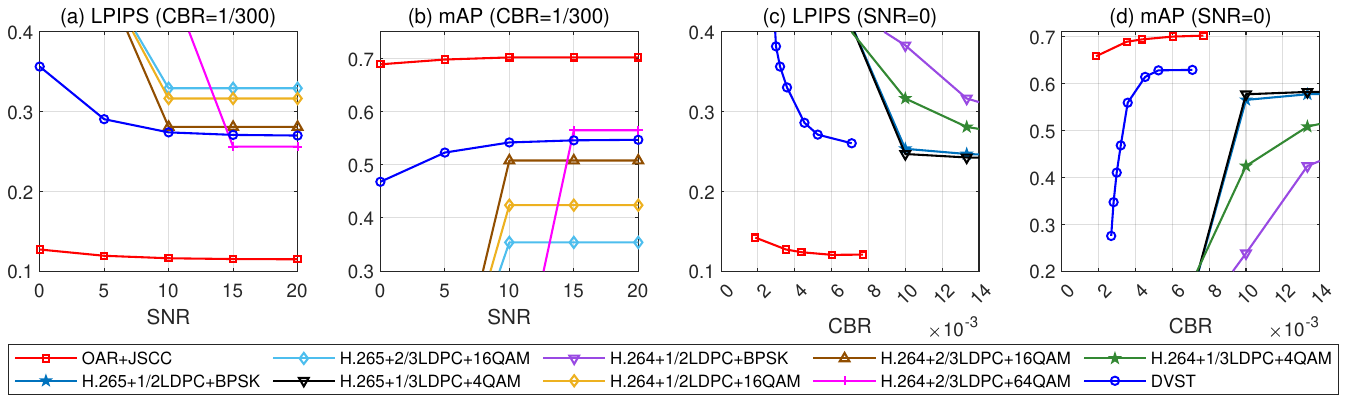}
  \caption{Performance comparison of the proposed OAR based method with H.264 and H.265 paired with LDPC and DVST on the CATER dataset.}
  \label{fig: video transmission_CATER}
\end{figure*}

\begin{figure*}
  \centering
  \includegraphics[width=1.0\textwidth]{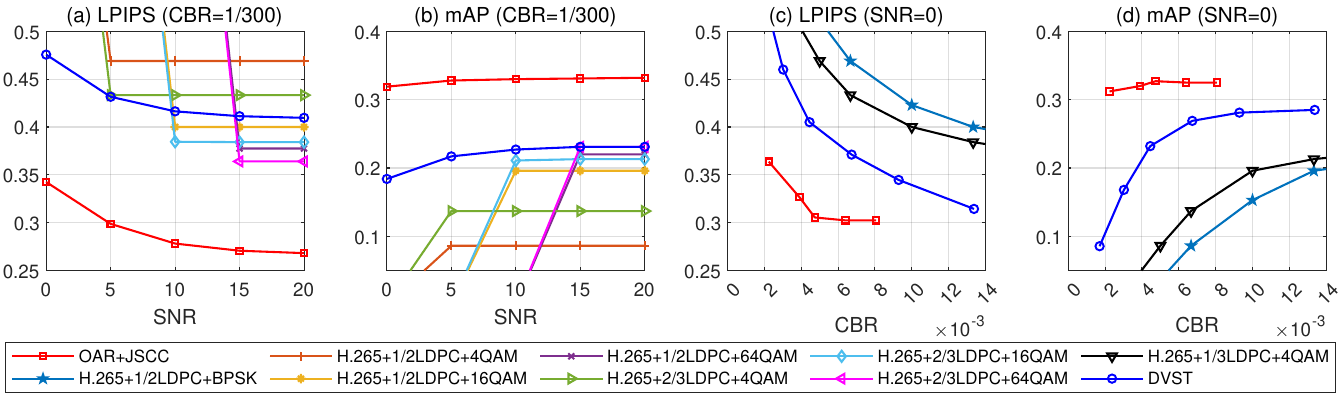}
  \caption{Performance comparison of the proposed OAR based method with H.265 paired with LDPC and DVST on the SoccerNet dataset.}
  \label{fig: video transmission_SoccerNet}
\end{figure*}

\subsection{Performance Evaluation of OAR-assisted JSCC}

\subsubsection{OAR-assisted Image JSCC for Reference Frame Transmission}

The performance of OAR-assisted JSCC is compared with BPG-based transmission and base JSCC for reference frame transmission on the UA-DETRAC dataset. Fig.~\ref{fig: LPIPS-SNR_CBR}(a)-(b) depict the image LPIPS variation with the SNR at different CBRs, and Fig.~\ref{fig: LPIPS-SNR_CBR}(c)-(d) illustrate the variation with the CBR at different SNRs. OAR-assisted JSCC generally outperforms base JSCC in general. Specifically, compared with most BPG+LDPC+QAM schemes, OAR-assisted JSCC avoids cliff effects. It also outperforms the BPG+1/3LDPC+4QAM and BPG+1/2LDPC+BPSK methods, which do not suffer from cliff effects in the 0$\sim$20 dB interval, particularly at high SNRs (Fig. \ref{fig: LPIPS-SNR_CBR}(b)). Furthermore, Fig.~\ref{fig: LPIPS-SNR_CBR}(c) demonstrates that OAR-assisted JSCC still maintains effective transmission at low SNRs, whereas most BPG-based transmissions encounter truncation, which fails to achieve effective transmission.

\subsubsection{OAR-assisted Video Transmission through JSCC}

The overall performance of the proposed video transmission system, which integrates OAR-based video generation and OAR-modulated JSCC, is assessed. Fig.~\ref{fig: CBR-SNR-Distortion} compares the performance of the proposed OAR-based video transmission system with traditional H.265-based mechanisms and the deep JSCC method DVST\cite{VideoJSCC} on the UA-DETRAC dataset.
Fig.~\ref{fig: CBR-SNR-Distortion}(a) and (b) demonstrate the variations in perceptual quality (LPIPS) and downstream task performance (mAP) at a CBR of 1/300. The proposed OAR-based method demonstrates superior perceptual quality and task performance, particularly at low SNRs. While H.265-based video transmission exhibits better quality at higher SNRs, it suffers from the cliff effect. Notably, under a CBR constraint of 1/300, the proposed method consistently outperforms DVST across all SNRs, achieving an average LPIPS loss reduction of 0.037 and an mAP enhancement of 0.16. 

Fig.~\ref{fig: CBR-SNR-Distortion}(c) and (d) illustrate the variation in performances with respect to the CBR at a fixed SNR of 0 dB. The proposed method achieves optimal perceptual quality and downstream task performance in the low CBR range (below 0.006). This is attributed to the ability to effectively reconstruct objects by leveraging the explicit semantics of OAR and prior knowledge of the generative model.
Compared with those of the other methods, the performance improvement of our proposed method is relatively gradual with increasing CBR. This is because other methods achieve video quality enhancement by transmitting residuals. Hence, our method is better suited for environments with severely limited bandwidth and significant noise interference.

Video transmission performances on the CATER and SoccerNet datasets are also evaluated from the same perspectives and presented in Fig.~\ref{fig: video transmission_CATER} and Fig.~\ref{fig: video transmission_SoccerNet}, respectively. The proposed OAR-based video transmission system maintains high perceptual quality and task performance even at low CBRs and low SNRs. This indicates the important role of explicit semantic information contained in OAR. Even though a low CBR and low SNR introduce a large distortion for the transmission of key frames, the video generator leverages the semantic prior to achieve faithful video synthesis based on OAR. This demonstrates the important advantage of the proposed framework for the transmission of semantically distinct scenarios involving objects with simple appearances.

\section{Analysis, Discussions and Ablation Studies}
\label{section 5}

\subsection{Ablation study on the deformation branch in OAR-based video generation}

To avoid redundancy, only the results of the experiments conducted on the UA-DETRAC dataset are presented below. The CATER and SoccerNet datasets yield similar conclusions. In the proposed video reconstruction framework shown in Fig.~\ref{fig: generation}, frame prediction is achieved via a two-branch fusion method: the synthesis branch (producing $\tilde{\bf I}_t$) and the deformation branch (producing $\check{\bf I}_t$). The final reconstructed image is obtained through a weighted summation of the outputs from the two branches. The deformation branch performs optical flow estimation based on the OAR layout of sequential frames and is used to synthesize images consistent with the motion characteristics. We quantitatively evaluate the impact of the deformation branch in terms of pixel-level distortion, perceptual distortion, and task performance. The results are presented in Table~\ref{tab: synthesis ablation}. 

\begin{table}[]
  \centering
  \caption{Results of the ablation study on the OAR-based video generation networks }
    \label{tab: synthesis ablation}
    \begin{tabular}{cccc}
    \toprule
                & PSNR  & LPIPS & mAP   \\ \midrule
    two branch ($\hat{\bf I}_t$)    & \textbf{19.73} & \textbf{0.389} & \textbf{0.565} \\
    two branch ($\tilde{\bf I}_t$)    & 19.33 & 0.401 & 0.540 \\
    single branch ($\hat{\bf I}_t$) & 18.62 & 0.438 & 0.498 \\
    \bottomrule
    \end{tabular}
\end{table}

The comparison between the synthesized image $\tilde{\bf I}_t$ and the final output image $\hat{\bf I}_t$ in the two-branch structure demonstrates that the introduction of the deformation branch significantly enhances the reconstruction performance. The reconstructed video after two-branch fusion shows noticeable performance gains compared with using only the synthesis branch. Furthermore, when the deformation branch is removed and the entire video reconstruction relies solely on the synthesis branch, substantial performance degradation is observed across all the metrics. This highlights the effectiveness of the deformation branch in improving video quality, particularly for long-term video synthesis.



\subsection{Appearance Enhancement with Semantic Priors}

\begin{figure}
  \centering
  \includegraphics[width=0.48\textwidth]{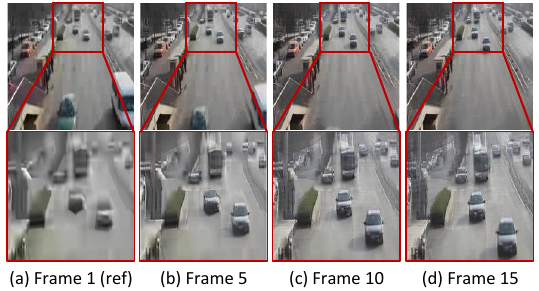}
  \caption{Video reconstruction results with reference frames of poor quality. The reference frame is taken as CRF 48.}
  \label{fig: warmup}
\end{figure}

Experiments conducted on OAR-based video coding at low bit-rates reveal that the proposed video generation model effectively enhances the reconstruction of objects via semantic priors. As illustrated in Fig.~\ref{fig: warmup}, this effect is demonstrated by showing the 5th, 10th, and 15th frames of the reconstructed video alongside their corresponding local details. The reconstruction model progressively incorporates prior information from the training set into the reconstructed frames, resulting in enhanced object details.

Accordingly, we propose a warm-up strategy to further validate the effectiveness of prior knowledge. Specifically, we take the compressed first frame as the reference frame and use the corresponding OAR for image generation. The above process is repeated for several iterations, with each generated image serving as the reference frame for the next iteration, which is called the warm-up stage. Subsequent frames are then generated after completing the warm-up stage. Table \ref{tab: warmup} presents the variations in mAP with warm-up of 10 iterations on the \texttt{val1} and \texttt{val2} datasets.
Warm-up at low bit-rates enhances the video quality to a certain extent. However, when the reference frames exhibit good quality (with CRF values of 42 and below), excessive inference by the model results in the loss of original details, leading to degradation of video quality.

\begin{table}
  \centering
    \caption{Variation in the mAP achieved by the warmup operation}
    \label{tab: warmup}
    \begin{tabular}{lllllll}
    \toprule
    \multirow{2}{*}{dataset} & \multicolumn{6}{c}{CRF of reference frame}                                         \\
                         & 48    & 45    & 42     & 39     & 36     & 33         \\ \midrule
    \texttt{val1}                 & {\bf 0.020} & {\bf 0.008} & -0.017 & -0.026 & -0.035 & -0.036 \\
    \texttt{val2}                 & {\bf 0.022} & {\bf 0.011} & -0.012  & -0.037  & -0.051 & -0.050 \\ \bottomrule
    \end{tabular}
    \end{table}
  
\subsection{Ablation Studies for OAR Attributes and Relations}

To assess the influence of each element in the OAR, ablation experiments are conducted across three dimensions: object categories, orientations, and relations. In each experiment, the same model is utilized for video generation, with different alterations to the OAR sequences: (1) w/o category: all categories are set to ``car"; (2) w/o angle: all angles to fixed to 0; (3) w/o relation: all occlusion relations are removed.
The performance variation in mAP is shown in Table~\ref{tab: ablation-overall}.
As indicated in the table, the object category and angle significantly affect the reconstruction performance, whereas the relation has a minimal impact. This suggests that occlusion relations may be implicitly embedded in spatial locations owing to the independence of objects, rendering them redundant in the OAR.
However, the inclusion of relations in OARs ensures the integrity of the graph structure, facilitating the adaptation and expansion of OAR representations to more complex scenes. 
Moreover, relations in OAR are able to convey high-level semantic information within a very small amount of data, which is conducive to the development of video transmission frameworks for both reconstruction and downstream tasks. Specifically, the receiver can perform downstream tasks either through direct reasoning based on the OAR or by using the OAR as auxiliary information for scene understanding. The relations in the OAR complement the behavioral logic and interactions between key entities in the scene at the semantic level, information that may not be effectively preserved in the reconstructed video under low bit-rate or low SNR conditions. An important direction for our future research is to explore how to further utilize the relations in OAR to achieve efficient video transmission and downstream task execution across a broader range of scenarios.

\begin{figure}
  \centering
  \includegraphics[width=1.0\linewidth]{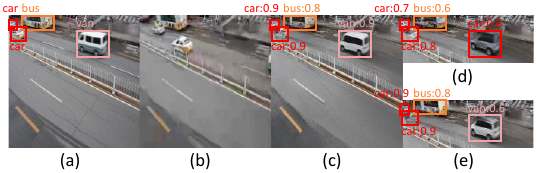}
  \caption{Visualization of ablation experiments. (a) Real image; (b) Reference frame; (c) Original reconstructed frame; (d) Reconstructed frame with the ablation setting: w/o category; (e) Reconstructed frame with the ablation setting: w/o angle. The bottom halves of (d) and (e) are the same as those in (c).}
  \label{fig: ablation}
\end{figure}

\begin{table}
  \centering
  \caption{Average variations in mAP metrics of videos obtained via different ablation methods}
  \label{tab: ablation-overall}
    \begin{tabular}{cccc}
      \toprule
      phase & w/o category & w/o angle & w/o relation \\
      \midrule
      \texttt{val1}  & -0.108       & -0.059    & -0.001       \\
      \texttt{val2}  & -0.105       & -0.052    & -0.001      \\
      \bottomrule
    \end{tabular}
  \end{table}

Fig.~\ref{fig: ablation} shows further video reconstruction results from ablation experiments (1) and (2).
As shown in Fig.~\ref{fig: ablation} (a) to (c), the van absent in the reference frame is generated leveraging category information from the OAR, and is correctly recognized as a ``van" with high confidence. 
This indicates the adaptability of OAR-based video transmission to downstream tasks. 
In Fig.~\ref{fig: ablation}(d), mislabeling the category as ``car" results in car-like appearances, such as a black body and a smaller front end.
However, the bus in the upper-left corner, which is influenced primarily by appearance information from the reference frame, remains unaffected by OAR mislabeling.
This reflects the ability of OAR to explicitly govern video semantics, suggesting its potential application in diverse domains, such as video editing and privacy protection. 
Fig.~\ref{fig: ablation}(e) shows the rectification of the model for angle mislabeling. This suggests that angle's impact on object appearance is not as pronounced as that of the category, manifesting instead through a composite effect with background or other features. On the other hand, this also indicates that the object orientation is also concurrently influenced by motion.

From the above analysis, it is evident that the attributes of object categories have explicit influence on appearance. To delve deeper into the interplay between categories, the impact of category mislabeling on various objects is examined.
Experiments show that the average precision of ``van" decreases by 0.316 and 0.340 on val1 and val2, respectively, whereas ``bus" and ``others" have relatively minor variations.

Compared with cars, objects categorized as ``bus" and ``others" are less susceptible to category mislabeling because of their distinct sizes and movement patterns.
Conversely, vans are most affected by mislabeling, with nearly half of the originally correctly recognized objects (with mean mAPs of 0.643 and 0.484 for \texttt{val1} and \texttt{val2}, respectively) being misidentified. Further experiments reveal that after being mislabeled as a car, the proportion of vans misidentified as cars increases from 35\% to 67\%. This finding indicates that objects with similar appearances are more susceptible to category labeling. This further exemplifies the explicit role of OAR in semantics and inspires awareness of the distinctiveness among different attributes during labeling design.

\begin{figure}[t]
  \centering
  \includegraphics[width=0.45\textwidth]{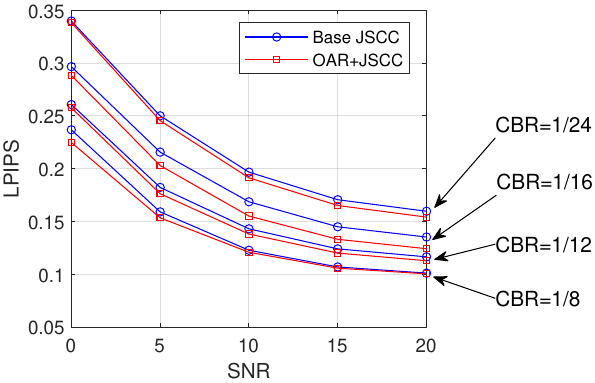}
  \caption{PSNR versus SNR curves of the foreground region obtained by different JSCC models for the three CBR values. 
  }
  \label{fig: OARvsBase}
\end{figure}

\subsection{Performance Variants for OAR-assisted JSCC}

Fig.~\ref{fig: OARvsBase} compares the LPIPS versus SNR performances of OAR-assisted JSCC with the base JSCC.
The experimental results indicate that the JSCC performance under different CBRs and SNRs is significantly improved by introducing OAR modulation, and an LPIPS loss reduction of up to 0.023 can be achieved. This finding indicates that the explicit semantic information contained in OAR plays an effective role in assisting the transmission and reconstruction of key frames. From the results of CBR=1/24 and 1/16, it is evident that the modulation effect of the OAR is dependent on a better channel environment. When transmitting with a lower CBR, a higher channel SNR is more helpful for the efficient expression of semantic information in OAR.
Since downstream tasks prioritize perceptual and semantic considerations, it is more valuable to optimize the loss function from perceptual and semantic perspectives \cite{GenerativeJSCC, PerceptualJSCC}. However, since this paper focuses primarily on video transmission with OAR as auxiliary information, refining the JSCC model specifically for reference frames falls outside the scope of this study.

\subsection{Inference time analysis}

The inference time of several key modules of the proposed framework was measured experimentally at two different image resolutions, and the results are presented in Table~\ref{tab: inference time}. The tested modules include JSCC encoding and decoding of key frames, OAR extraction, OAR embedding and graph computation, and image generation. The experiments were conducted on an A800 GPU. It should be noted that the current system focuses on the proposal of the OAR-based video transmission system and its rate-semantic-distortion performance, while computational complexity is not the primary focus.

\begin{table}[]
  \centering
  \caption{Inference time for different modules}
    \label{tab: inference time}
  \begin{tabular}{ccc}
  \toprule
  module                            & \begin{tabular}[c]{@{}c@{}}inference time/ms\\ ($512\times512$)\end{tabular} & \begin{tabular}[c]{@{}c@{}}inference time/ms\\ ($256\times256$)\end{tabular} \\ \midrule
  JSCC Encoding                      & 31.20                    & 14.72                                                              \\
  JSCC Decoding                      & 6.92                     & 7.44                                                               \\
  OAR Extraction                    & 9.8                      & 1.4                                                                \\
  \begin{tabular}[c]{@{}c@{}}OAR Embedding\\ and Graph Computing\end{tabular} & 18.80                    & 9.22                                                               \\
  Image Generation                  & 22.35                    & 9.18                                                               \\
  Total                             & 89.07                    & 41.96                                                              \\ \bottomrule
  \end{tabular}
\end{table}

However, various optimization techniques can enhance inference speed without significantly compromising video quality. In terms of network design, smaller models can be achieved by reducing the number of convolutional kernels, convolutional layers, and embedding dimensions, thereby accelerating model inference. Additionally, techniques such as knowledge distillation and pruning can be employed to create lightweight models. Model efficiency and inference speed can also be improved through parameter quantization and low-precision inference. Furthermore, pipeline design and parallel processing of multiple frames offer potential acceleration methods. However, these optimizations fall outside the scope of this paper and will be addressed in our future research.

\section{Conclusion}
\label{section 6}

In this paper, an OAR-assisted video coding and transmission method was proposed to address the challenge of efficient and reliable video transmission for downstream tasks. First, an OAR-based video representation and coding system was proposed to realize a low bit-rate representation of videos via object-attribute-relation (OAR). Second, a generative video reconstruction method based on reference frames and OAR sequences was implemented to realize perceptual-quality and downstream-task oriented reconstruction. Finally, OAR-assisted JSCC was proposed and combined with OAR-based video coding to realize video transmission in noisy channels. Experiments on a traffic surveillance dataset demonstrated the effectiveness of our approach. First, the proposed OAR-based video coding outperformed H.265 in terms of perceptual quality and downstream object detection performance and saved up to 61\% bit-rate. Second, the implemented OAR-based video transmission system could be adaptable to low bandwidth and SNR conditions, effectively ensuring perceptual quality and downstream task performance. At a CBR of 1/300, our method achieved an average LPIPS loss reduction of 0.054 and an improvement in the object detection performance (mAP) of 0.15 compared with DVST. Additionally, the scalability of the proposed OAR representation suggests its potential applicability in complex scenarios such as border and sea defenses, as well as in applications requiring explicit semantic control such as privacy-preserving video transmission.

\bibliographystyle{IEEEtran}
\normalem
\bibliography{IEEEabrv, main}

\clearpage

\begin{IEEEbiography}[{\includegraphics[width=1in,height=1.25in,clip,keepaspectratio]{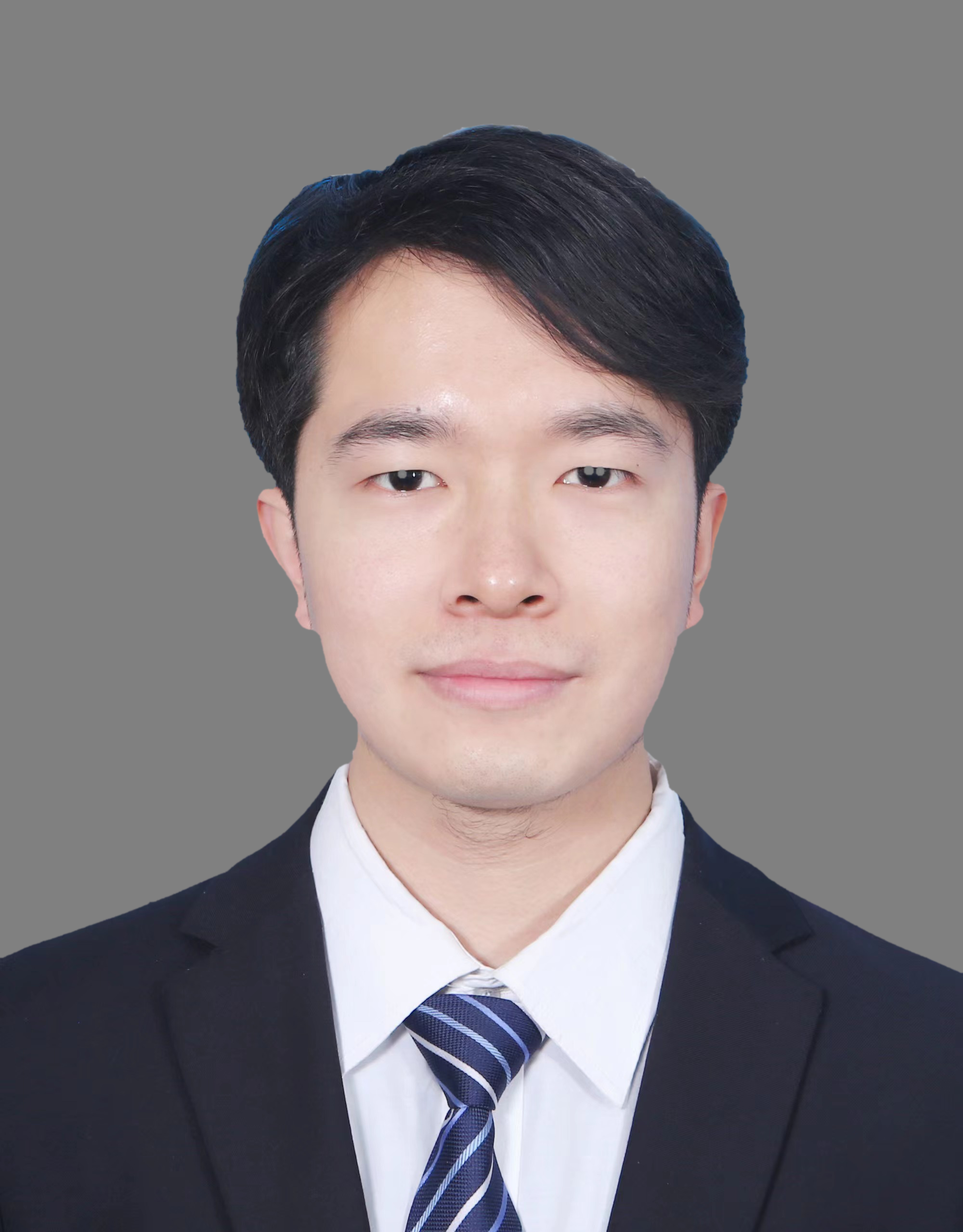}}]{Qiyuan Du}
  received the B.E. degree from the department of Electronic Engineering, Tsinghua University, Beijing, China, in 2021. He is currently pursuing the Ph.D. degree. His current research interests include machine learning, computer vision, multimedia communications, and semantic communication.
\end{IEEEbiography}

\begin{IEEEbiography}[{\includegraphics[width=1in,height=1.25in,clip,keepaspectratio]{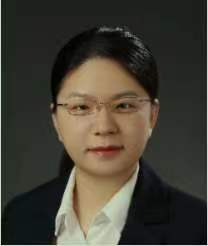}}]{Yiping Duan}
  received a B.S. degree from the School of Computer Science and Technology, Henan Normal University, Xinxiang, China, in 2010, and a Ph.D. degree from the School of Computer Science and Technology, Xidian University, Xian, China, in 2016.

  She is currently an Assistant Research Fellow with the Department of Electronic Engineering, Tsinghua University, Beijing, China. Her current research interests include wireless communications, machine learning, computer vision, and image processing.
\end{IEEEbiography}

\begin{IEEEbiography}[{\includegraphics[width=1in,height=1.25in,clip,keepaspectratio]{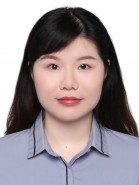}}]{Qianqian Yang} received the B.Sc. degree in automation from Chongqing University, Chongqing, China, in 2011, the M.Sc. degree in control engineering from Zhejiang University, Hangzhou, China, in 2014, and the Ph.D. degree in electrical and electronic engineering from Imperial College London, London, U.K. She has held visiting positions with CentraleSupelec, Gif-sur-Yvette, France, in 2016, and the New York University Tandon School of Engineering, Brooklyn, NY, USA, from 2017 to 2018. After her Ph.D., she worked as a Postdoctoral Research Associate with Imperial College London and as a Machine Learning Researcher with Sensyne Health Plc, Oxford, U.K. She is currently a Tenure-Tracked Professor with the Department of Information Science and Electronic Engineering, Zhejiang University.
\end{IEEEbiography}

\begin{IEEEbiography}[{\includegraphics[width=1in,height=1.25in,clip,keepaspectratio]{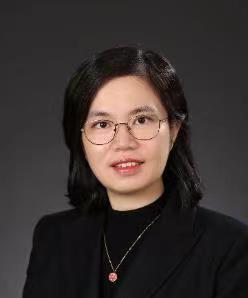}}]{Xiaoming Tao}
  is currently a Full Professor at the Department of Electronic Engineering, Tsinghua University. Her research focuses on semantic coding and computing communications for multimedia.  In the related areas, she has published over 60 journal and conference papers in addition to over 40 granted patents.

  Dr. Tao was a recipient of the National Science Foundation for Outstanding Youth, from 2017 to 2019, and many national awards, e.g., the 2017 China Young Women Scientists Award, the 2017 Top Ten Outstanding Scientists and Technologists from the China Institute of Electronics, the 2017 First Prize of the Wu Wen Jun A.I. Science and Technology Award, the 2016 National Award for Technological Invention Progress, and the 2015 Science and Technology Award of the China Institute of Communications. She served as the workshop general co-chair for the IEEE INFOCOM 2015, the organization co-chair for the IEEE ICCI*CC 2015/2020, and the volunteer leader for IEEE ICIP 2017. She is currently an editor of IEEE Transactions on Wireless Communications, China Communications, and Pattern Recognition, as well as the scientific editor of Chinese Journal of Electronics.
\end{IEEEbiography}

\begin{IEEEbiography}[{\includegraphics[width=1in,height=1.25in,clip,keepaspectratio]{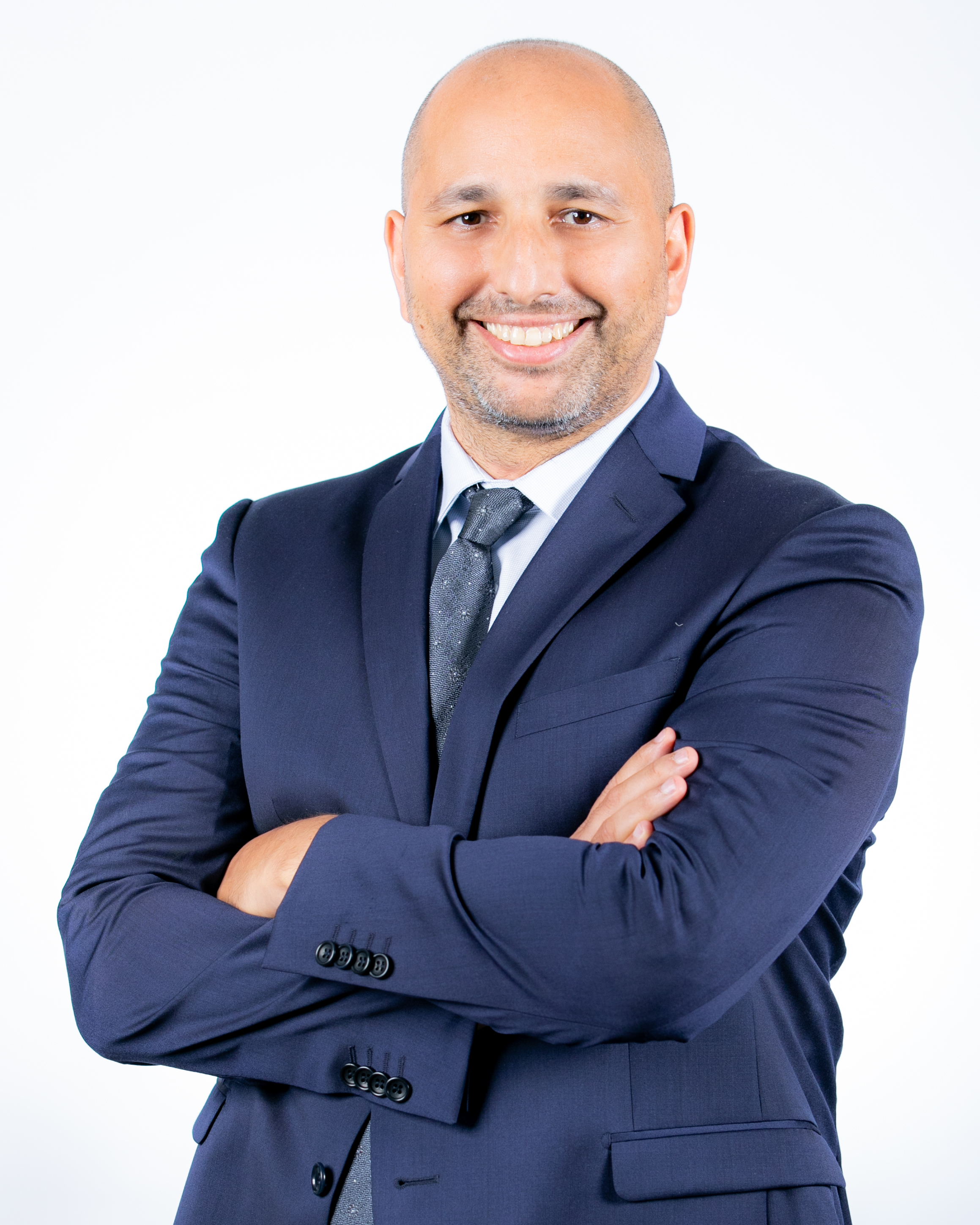}}]{M\'{e}rouane Debbah} is a Professor at Khalifa University of Science and Technology in Abu Dhabi and founding Director of the KU 6G Research Center. He is a frequent keynote speaker at international events in the field of telecommunication and AI. His research has been lying at the interface of fundamental mathematics, algorithms, statistics, information and communication sciences with a special focus on random matrix theory and learning algorithms. In the Communication field, he has been at the heart of the development of small cells (4G), Massive MIMO (5G) and Large Intelligent Surfaces (6G) technologies. In the AI field, he is known for his work on Large Language Models, distributed AI systems for networks and semantic communications. He received multiple prestigious distinctions, prizes and best paper awards (more than 40 IEEE best paper awards) for his contributions to both fields. He is an IEEE Fellow, a WWRF Fellow, a Eurasip Fellow, an AAIA Fellow, an Institut Louis Bachelier Fellow, an AIIA Fellow and a Membre \'{e}m\'{e}rite SEE. He is actually chair of the IEEE Large Generative AI Models in Telecom (GenAINet) Emerging Technology Initiative and a member of the Marconi Prize Selection Advisory Committee.
\end{IEEEbiography}

\end{document}